\documentclass[prl,twocolumn,preprintnumbers,amsmath,amssymb,nofootinbib,epsfig,bmamsfonts,yfonts,superscriptaddress]{revtex4}
\usepackage{graphicx}

\relax

\begin{document}

\title{Accelerating cosmological expansion from shear and bulk viscosity}


\author{Stefan Floerchinger}
\affiliation{Physics Department, Theory Unit, CERN, CH-1211 Gen\`eve 23, Switzerland}
\author{Nikolaos Tetradis}
\affiliation{Department of Physics, University of Athens, Zographou 157 84, Greece}
\affiliation{Physics Department, Theory Unit, CERN, CH-1211 Gen\`eve 23, Switzerland}
\author{Urs Achim Wiedemann}
\affiliation{Physics Department, Theory Unit, CERN, CH-1211 Gen\`eve 23, Switzerland}



\begin{abstract} 
The dissipation of energy from local velocity perturbations in the cosmological fluid affects the time evolution of spatially averaged fluid dynamic fields and the cosmological solution of Einstein's field equations. We show how this backreaction effect depends on shear and bulk viscosity and other material properties of the dark sector, as well as the spectrum of perturbations. If sufficiently large, this effect could account for the acceleration of the cosmological expansion.
\end{abstract}

\maketitle

The possibility that dissipative phenomena in the form of bulk viscosity might affect the expansion history of the universe has been discussed repeatedly. Indeed, for a homogeneous and isotropic expansion, the bulk viscous pressure is negative and could account for effects usually attributed to dark energy and held responsible for the current accelerating expansion of the universe~\cite{Murphy:1973zz,Padmanabhan:1987dg,Fabris:2005ts}. However, models that aim at replacing the need for separate dark matter and dark energy components in the cosmological concordance model by invoking bulk viscosity of dark matter are strongly challenged by cosmological precision data~\cite{Li:2009mf,Gagnon:2011id,Velten:2011bg,Velten:2012uv}. In this letter, we point out that dissipative effects relevant for the expansion history of the universe could arise also from the shear viscous properties of the cosmological fluid. Also, at least in principle, one could have a similar effect from the gain in internal energy due to fluid motion against local pressure gradients. 

In general relativity, the matter fields that enter the energy-momentum tensor $T_{\mu\nu}$, as well as the metric $g_{\mu\nu}$ that enters $T_{\mu\nu}$ and the Einstein tensor $G_{\mu\nu}$, are dynamical variables. The time evolution of the latter is determined by Einstein's field equations
\begin{equation}
	G_{\mu\nu} = -8\pi G_\text{N}\, T_{\mu\nu}\, .
	\label{eq1}
\end{equation}
If the universe were completely homogeneous and isotropic, 
%
%
the energy-momentum tensor could deviate from its ideal form at most by a bulk viscous term, $T^0_{\;\;0} =\epsilon$, $T^i_{\;\,j}=(p + \pi_\text{bulk}) \delta^i_{\;j}$. Einstein's equations would then reduce to the standard Friedmann equations that express the time evolution of the scale factor $a(\tau)$ in terms of the energy density $\epsilon$ and the effective pressure $p_\text{eff} = p+\pi_\text{bulk}$.

For the more realistic case of a universe that is homogeneous and isotropic only in a statistical sense, a  Friedmann-type solution acts as a background. The evolution of this background is not affected by the perturbations if they are small enough for only linear terms to be kept. However, the fluctuations can backreact on the background at non-linear order. On the gravity side (left hand side of eq.\ \eqref{eq1}, broadly speaking), these backreaction effects have come under scrutiny and are likely to be small \cite{Wetterich:2001kr,Green:2010qy}. Here, we discuss backreaction effects on the matter side of eq.\ \eqref{eq1}. 

For matter described as a relativistic viscous fluid,  the energy-momentum tensor in the Landau frame (where the fluid velocity is defined by the condition that there is no energy current
in the fluid rest frame, $-u_\mu T^{\mu\nu}=\epsilon u^\nu$) reads
\begin{equation}
	T^{\mu\nu} = \epsilon \, u^\mu u^\nu + (p + \pi_\text{bulk}) \Delta^{\mu\nu} + \pi^{\mu\nu}.
\label{eq3}
\end{equation}
Here, $\Delta^{\mu\nu} = u^\mu u^\nu + g^{\mu\nu}$ is a projector orthogonal to the fluid velocity, and $\pi^{\mu\nu}$ is the shear stress, satisfying $u_\mu \pi^{\mu\nu} = \pi^\mu_{\;\;\mu}=0$. 
To first order in the gradient expansion of hydrodynamics, one has the following constitutive relations
\begin{eqnarray}
\pi^{\mu\nu} & = & - 2 \, \eta \, \sigma^{\mu\nu} \label{eq4} \\
& = & - \eta  \left[ \Delta^{\mu\alpha} \Delta^{\nu\beta} +  \Delta^{\mu\beta} \Delta^{\nu\alpha} - \frac{2}{3} \Delta^{\mu\nu} \Delta^{\alpha\beta} \right] \nabla_\alpha u_\beta\, , \nonumber
\\
\pi_\text{bulk} & = & - \zeta \, \Theta  = - \zeta \, \nabla_\mu u^\mu\, ,
\label{eq5}
\end{eqnarray}
where $\eta$ and $\zeta$ denote the shear and bulk viscosity, respectively. In addition, there can be conserved charges. 
For a single conserved current $N^{\alpha}$ (corresponding e.g. to conserved baryon number or a conserved number of dark matter particles), 
one has to first order in hydrodynamical gradients a particle diffusion current $\nu^{\alpha}$ that points along chemical potential gradients 
orthogonal to the fluid velocity. Its strength is set by the thermal conductivity $\kappa$:
\begin{eqnarray}
N^\alpha &=& n \, u^\alpha + \nu^\alpha\, ,\label{eq6}\\
\nu^\alpha &=& - \kappa \left[ \frac{n T}{\epsilon + p} \right]^2 \Delta^{\alpha\beta}\partial_{\beta}\left(\frac{\mu}{T}\right)\, .
\label{eq7}
\end{eqnarray}
While keeping the next (second) order in the gradient expansion is important for maintaining causal dynamics and linear stability~\cite{Israel:1979wp,Hiscock:1985zz}, first-order viscous hydrodynamics is usually sufficient for practical calculations, and it reduces in the non-relativistic limit to the conventional Navier-Stokes theory. We therefore restrict the following discussion to it. 

From the covariant conservation of energy, momentum and particle number, 
\begin{equation}
\nabla_\mu T^{\mu\nu} = 0, \quad\quad \nabla_\mu N^\mu = 0\, ,
\label{eq8}
\end{equation}
one finds the fluid dynamic equations of motion for the energy density 
\begin{equation}
u^\mu \partial_\mu \epsilon + (\epsilon + p ) \nabla_\mu u^\mu   - \zeta \Theta^2 - 2 \eta \sigma^{\mu\nu} \sigma_{\mu\nu} = 0\, ,
\label{eq9}
\end{equation}
the fluid velocity
\begin{equation}
(\epsilon + p + \pi_\text{bulk}) \, u^\mu \nabla_\mu u^\nu  + \Delta^{\nu\mu} \, \partial_\mu (p + \pi_\text{bulk}) +\Delta^\nu\,_\alpha \nabla_\mu \pi^{\mu\alpha}= 0\, ,
\label{eq10}
\end{equation}
and the particle number density
\begin{equation}
u^\mu \partial_\mu n +n \nabla_\mu u^\mu +\nabla_\mu \nu^\mu =0 \, .
\label{eq11}
\end{equation}
In eqs.~\eqref{eq9}, \eqref{eq10} and \eqref{eq11}, only energy density $\epsilon = u_\mu u_\nu T^{\mu\nu}$ and particle number density $n = u_\mu N^\mu$ are independent thermodynamic variables. 

Einstein's field equations (\ref{eq1}) imply the conservation laws (\ref{eq8}) and thus the equations of motion~\eqref{eq9}, \eqref{eq10} and \eqref{eq11}. Here we work with the latter. Once supplemented by an e.o.s.,\ they form a closed set for the evolution of fluid dynamic fields. At least locally, these equations provide sufficient information about the time evolution of the thermodynamic variables and the fluid velocity. Also, eqs.\ \eqref{eq9}, \eqref{eq10} and \eqref{eq11} are valid for arbitrary gravitational fields, on which they depend via the dependence of the fluid velocity $u^\mu$, the projector $\Delta^{\mu\nu}$ and the covariant derivatives $\nabla_\mu$ on the metric $g_{\mu\nu}$. 
To analyze this dependence in more detail, let us now consider a perturbative ansatz for the metric 
\begin{equation}
ds^2=a^2(\tau)\left[
-\left(1+2\Psi(\tau,\vec x) \right)d\tau^2
+\left(1-2\Phi(\tau,\vec x) \right) d \vec x \cdot d\vec x \right],
\label{eq12} 
\end{equation}
where $\Phi$ and $\Psi$ denote potentials (in conformal Newtonian gauge) and $a(\tau)$ is the scale factor.   We follow here the general expectation that, at least at late times, the main modification of a simple homogeneous and isotropic expansion is mediated by scalar fluctuations around the background metric, and that the influence of vector and tensor excitations is negligible \cite{Weinberg:2008zzc}. 

With the metric of eq.\ \eqref{eq12}, the fluid velocity can be written as $u^\mu =\left(\gamma, \gamma \vec v\right)$, where $\gamma = 1/(a\sqrt{1-\vec v^2+2\Psi+2\Phi \vec v^2})$ (in units where $c=1$). We specialize now to the cosmologically relevant case of small fluid velocity, $\vec v^2 \ll 1$. The different terms entering eqs.\ \eqref{eq9}, \eqref{eq10} and \eqref{eq11} can be computed, and to linear order in $\Phi, \Psi$, one finds for instance
\begin{equation}
\begin{split}
\nabla_\mu u^{\mu} = & \tfrac{1}{a} {\big [} \vec{\nabla} \cdot \vec{v} + 3 \tfrac{\dot a}{a} \\
& - \Psi\vec{\nabla} \cdot \vec{v} - 3\tfrac{\dot a}{a} \Psi - 3 \dot \Phi -3 \vec{v} \cdot \vec{\nabla}\Phi  {\big ]} .
\end{split}
\label{eq13}
\end{equation}
In the regime of structure formation at late times, one expects that the Newton potentials are small ($\Phi\, ,\Psi \ll 1$), that they vary slowly in time (typically with the Hubble rate, $\dot\Phi \sim \tfrac{\dot a}{a} \Phi$, and similarly for $\Psi$) and that they vary in space on similar scales as the fluid dynamic fields~\cite{Weinberg:2008zzc}. In this case, only the first two terms on the right hand side of eq.\ \eqref{eq13}, i.\ e.\ the ones that are independent of $\Phi$ and $\Psi$, must be kept. We analyze other terms in eq.\ \eqref{eq9} in a similar way and find that it becomes
\begin{equation}
\begin{split}
&\dot \epsilon + \vec v \cdot \vec \nabla \epsilon +  (\epsilon + p) \left(3 \tfrac{\dot a}{a} + \vec \nabla \cdot \vec v \right) \\
& = \tfrac{\zeta}{a} \left[3\tfrac{\dot a}{a} + \vec \nabla \cdot \vec v\right]^2 + \tfrac{\eta}{a} \left[ \partial_i v_j \partial_i v_j + \partial_i v_j \partial_j v_i - \tfrac{2}{3} (\vec \nabla \cdot \vec v)^2 \right] ,
\end{split}
\label{eq14}
\end{equation}
where sub-leading $\Phi$- and $\Psi$-dependent terms have now been suppressed. An analogous argument~\footnote{The situation is different for the time evolution of the fluid velocity \eqref{eq10}, where scalar fluctuations in the metric enter to leading order. 
For instance, eq.~\eqref{eq10} contains the Newtonian acceleration in a term $\dot v_i + \vec v\cdot \vec \nabla v_i + \tfrac{\dot a}{a} v_i + \partial_i \Psi$. In this letter 
we do not use the evolution of the fluid velocities.}
applies also to the time evolution (\ref{eq11}) of the particle number density that reads in the same limit 
\begin{equation}
\dot n + \vec v \cdot \vec \nabla n + n \left(3 \tfrac{\dot a}{a} + \vec \nabla \cdot \vec v \right) 
 = \tfrac{1}{a} \vec \nabla \cdot \left[ \kappa \left( \tfrac{nT}{\epsilon + p} \right)^2 \vec \nabla \left(\tfrac{\mu}{T}\right) \right] .
 \label{eq15}
\end{equation}

We turn next to the expectation values or spatial averages $\bar \epsilon = \langle \epsilon \rangle$ and $\bar n = \langle n \rangle$. From eq.\ \eqref{eq15}, one finds (neglecting surface terms as usual)
\begin{equation}
\tfrac{1}{a} \dot{\bar n} + 3 H \, \bar n = 0,
\label{eq16}
\end{equation}
with Hubble parameter $H=\dot a / a^2$. This shows simply that the standard dilution of particle number due to the expansion
is not modified by dissipative effects. On the other hand, we find from eq.\ \eqref{eq14} for the cosmological evolution  of the average energy density
\begin{equation}
\tfrac{1}{a}\dot{\bar \epsilon} + 3 H \, (\bar \epsilon + \bar p - 3 \bar \zeta H) =D\, ,
\label{eq17}
\end{equation}
where we have introduced the shorthand
\begin{equation}
\begin{split}
& D=  \tfrac{1}{a^2} \langle \eta \left[ \partial_i v_j \partial_i v_j + \partial_i v_j \partial_j v_i - \tfrac{2}{3} \partial_i v_i \partial_j v_j\right] \rangle\\
& \quad\quad + \tfrac{1}{a^2} \langle \zeta[\vec \nabla \cdot \vec v]^2 \rangle +
\tfrac{1}{a}\langle  \vec v \cdot \vec \nabla \left( p - 6 \zeta H \right)\rangle \, .
\end{split}
\label{eq18}
\end{equation}
The term $D$ enters eq.\ \eqref{eq17} as a backreaction of fluid fluctuations onto the time evolution of the background field $\bar \epsilon$. The shear and bulk viscous contributions of the first and second term on the r.h.s. of eq.~\eqref{eq18} are positive semi-definite, since they are expectation values of squares. They describe the gain in internal energy due to the dissipation of local gradients in the fluid velocity. The third term accounts for the work done by the fluid expanding out of high-pressure regions (or a corresponding gain in internal energy due to contraction against a pressure gradient). 

When structures form through local gravitational collapse, one expects that the shear and bulk viscous contributions to $D$ increase the internal energy. Depending on the equation of state and the dissipative properties of the fluid, this effect might be small or sizable. In the following we simply assume that it is non-negligible and discuss possible consequences for the cosmic expansion.

To do so, we need to supplement the fluid dynamic evolution equations with an equation for the scale factor. A direct spatial average of Einstein's field equation \eqref{eq1} with the energy-momentum tensor \eqref{eq3} would involve on its right hand side unknown quantities such as $\langle (\epsilon + p + \pi_\text{bulk}) u^\mu u^\nu \rangle$. One could project to the different terms in eq.\ \eqref{eq3} by contracting with the fluid velocity, e.\ g.\ $u^\mu u^\nu G_{\mu\nu} = - 8 \pi G_\text{N} \epsilon$, but the space average of this equation would involve unknown averages of velocities on the left hand side. We, therefore, look in Einstein's equations for a suitable constraint that is independent of $u^\mu$, and find it in the trace $R=8\pi G_\text{N}\, T^\mu_{\;\;\mu}$.
The averaged part $\langle R \rangle =8\pi G_\text{N}\, \langle T^\mu_{\;\;\mu}\rangle$ reads
\begin{equation}
\frac{\ddot a}{a^3} = \frac{1}{a} \dot H + 2 H^2 = \frac{4\pi G_\text{N}}{3} \left(\bar \epsilon - 3 \bar p - 3 \bar \pi_\text{bulk}\right).
\label{eq19}
\end{equation}
For given e.o.s.\ and thermodynamic transport properties, one can determine the time evolution of the Hubble parameter and the scale factor $a(\tau)$ by solving eq.~\eqref{eq19} together with eqs.\ \eqref{eq16} and \eqref{eq17}. However, one also needs the parameter $D$ in eq.\ \eqref{eq18}, which depends on correlation functions of perturbations.

\begin{figure}
\includegraphics[width=0.35\textwidth]{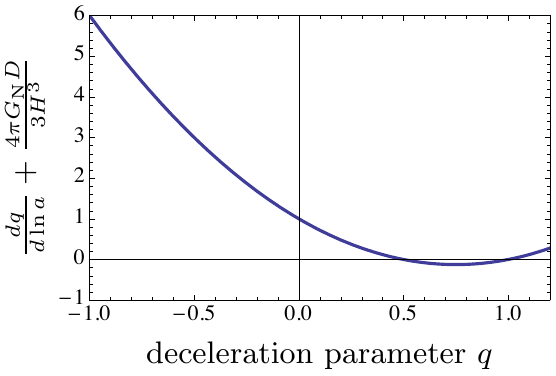}
\caption{Graphical representation of the evolution equation \eqref{eq20} for the deceleration parameter $q$ for the case of vanishing effective pressure, $\hat w=0$.}
\label{fig1}
\end{figure}

To illustrate the physics encoded in this set of equations, we assume first for the e.o.s.\ a simple relation $\bar p+\bar \pi_\text{bulk} = \hat w \, \bar \epsilon$, with $\hat w$ a numerical constant. A straightforward calculation gives for the deceleration parameter $q=-1-\dot H/ (a H^2)$
\begin{equation}
-\frac{d q}{d \ln a} + 2 (q-1) \left(q-\tfrac{1}{2}(1+3 \hat w)\right) = \frac{4\pi G_\text{N} D(1-3\hat w)}{3 H^3}.
\label{eq20}
\end{equation}
For $D=0$, eq.\ \eqref{eq20} has an attractive fixed point at the well known value $q=(1+3\hat w)/2$. In particular, for a pure cold dark matter universe with $\hat w = 0$ and with negligible dissipation, $D=0$, one finds deceleration, $q = 1/2$.
Interestingly, if the right hand side of equation \eqref{eq20} is positive, the fixed point is shifted to more negative values of $q$. 
More specifically, the fixed point is accelerating, i.\ e.\ $q<0$, for
\begin{equation}
\frac{4\pi G_\text{N} D}{3 H^3} > \frac{1+3\hat w}{1-3\hat w}\, .
\label{eq21}
\end{equation}
As a result, a positive $D$ can actually contribute to the acceleration of the expansion, similarly to an effective negative pressure $\hat w<0$ induced by bulk viscous pressure $\bar\pi_\text{bulk}$, or a positive cosmological constant, or dark energy. Fig.\ \ref{fig1} illustrates eq.\ \eqref{eq20} graphically and shows in particular the value the dissipative term must take in order to account for a given deceleration parameter $q$. We concentrate here on vanishing effective pressure, $p_\text{eff}=0$. For the experimentally favored value of $q\approx - 0.6$~\cite{Ade:2013zuv}, we conclude that the set of equations \eqref{eq17}, \eqref{eq19} could account for the observed accelerating expansion of the universe if $4\pi G_\text{N} D/(3 H^3)\approx 3.5$ (assuming $|dq/d\ln a | \ll 1$).

The above analysis of the acceleration parameter was done for a very simple e.o.s.,  $\bar p+\bar \pi_\text{bulk} = \hat w \, \bar \epsilon$.
For a universe filled with pure radiation,  $\hat w \to 1/3$, no finite dissipative term $D$ can satisfy eq.~\eqref{eq21}. But for a more realistic e.o.s., 
one has $p = p( \epsilon,n)$, and eq.~\eqref{eq20} for the acceleration parameter is replaced by a more complicated expression that
depends on first and second derivatives of the thermodynamic potential  $p( \epsilon,n)$. One has to check then for each e.o.s. whether the
combination of eqs.\ \eqref{eq16}, \eqref{eq17} and \eqref{eq19} contributes to deceleration or acceleration, and how sizable the effect can 
be. 

It is useful to decompose $\vec v $ in eq.\ \eqref{eq18} into a sum of a gradient, characterized by $\theta = \vec \nabla \cdot \vec v$, and a rotation, characterized by the vorticity $\vec w= \vec\nabla \times \vec v$, and to go to Fourier space, $\theta(x) = \int d^3 q \, \tilde \theta(q) \, e^{i\vec q \vec x}$ etc., 
\begin{equation}
\begin{split}
D  = & - \tfrac{1}{a} \int d^3 q \, P_{\theta p}(\vec q) + \tfrac{1}{a^2} \left(\bar \zeta + \tfrac{4}{3}\bar \eta \right) \int d^3 q \, P_{\theta\theta} (\vec q) \\
& + \tfrac{1}{a^2} \bar \eta \int d^3 q  \, (P_w)_{jj}(\vec q)\, .
\end{split}
\label{eq22}
\end{equation}
We assumed for simplicity that $\zeta = \bar \zeta$ and $\eta=\bar \eta$ are constant in space, and we defined  
the power spectra 
\begin{equation}
\begin{split}
\label{eq23}
\langle \tilde \theta(\vec q_1) \tilde p(\vec q_2) \rangle & = \delta^{(3)}(\vec q_1 + \vec q_2) P_{\theta p}(\vec q_1),\\
\langle \tilde\theta(\vec q_1) \tilde\theta(\vec q_2) \rangle & = \delta^{(3)}(\vec q_1 + \vec q_2) P_{\theta\theta}(\vec q_1),\\
\langle \tilde w_i(\vec q_1) \tilde w_j(\vec q_2) \rangle & = \delta^{(3)}(\vec q_1 + \vec q_2) (P_w)_{ij}(\vec q_1)\, .
\end{split}
\end{equation}
If the spectra $P_{\theta p}(\vec q)$, $P_{\theta\theta} (\vec q)$  and $(P_w)_{jj}(\vec q)$ in eq.~\eqref{eq22} do not die out faster than
$1/q^3$ ,  $D$ is dominated by the UV, i.\ e.\ by the fine structures in position space. 
Hence, one expects that the value of $D$ will be set by the smallest relevant scale. This is the dissipation or virialization scale, below which a fluid dynamic description does not apply.\footnote{In a companion paper~\cite{BFTW2014} we have analyzed in a technically more detailed way a backreaction effect that arises in a simple fluid dynamic model of heavy ion collisions and is analogous to eqs.\ \eqref{eq17} and \eqref{eq18}.}

Leaving a detailed study of $D$ to future work, we explore here the possibility that it could be sizeable, in the sense that $4\pi G_\text{N} D /(3H^3) \gtrsim 1$ and eq.\ \eqref{eq20} allows for accelerating expansion.
 It is generally difficult to conceive that bulk viscosity is large enough to have a substantial effect, in particular because neither radiation (ultra-relativistic particles) nor simple non-relativistic gases can contribute to it~\cite{Weinberg:1971mx} (see, however, ref.\ \cite{Gagnon:2011id} for a counterexample). 
We therefore focus on the shear viscous part of $D$. We simply assume that typical gradients of the fluid velocity are of the same order as the Hubble rate $H$, so that 
$\bar\eta \langle \partial_i v_j \partial_i v_j + \partial_i v_j \partial_j v_i - \tfrac{2}{3} \partial_i v_i \partial_j v_j\rangle/ a^2 = \sigma \bar \eta H^2$ 
with $\sigma$ of order one. This corresponds to realistic peculiar velocity variations of the order of 100 $\text{km}/s$ on distances of $1\, \text{MPc}$. It amounts essentially to assuming that
the smallest distances relevant for $D$ are of this order.

It remains to estimate the shear viscosity. In general, this will depend on the unknown material properties of the dark sector.
It is noteworthy, however, that a large shear viscosity arises for systems containing very weakly interacting relativistic particles of long mean free paths (e.g. forming an additional component 
to cold dark matter with shorter range interactions)~\cite{Weinberg:1971mx}. In this case,
relativistic kinetic theory suggests that~\cite{Weinberg:2008zzc} 
\begin{equation}
\eta =
c_\eta
\epsilon_R \tau_R\, ,
\label{eq24}
\end{equation}
where $c_\eta $ is a numerical prefactor of order one, $\epsilon_R$ is the energy density carried by the weakly interacting particles and $\tau_R$ is their mean free time. Accelerating expansion would result if the e.o.s. of this system is not pure radiation and if 
\begin{equation}
\frac{4\pi G_\text{N} D}{3 H^3} = \frac{c_\eta \epsilon_R \tau_R H \sigma}{2 \rho_c}
\label{eq25}
\end{equation}
is of order unity, where the critical energy density $\rho_c$ is defined by $H^2=8\pi G_\text{N} \rho_c / 3$. On the other hand, for a description in terms of a single fluid to be applicable, the mean free times of the weakly interacting particles must be smaller than the expansion rate, $\tau_R H<1$. Thus, the term in \eqref{eq25} can only become of order one if $\epsilon_R$ is of the same order as $\rho_c$ and if $\sigma$ is somewhat larger than one. 

One may wonder whether there is any reasonable weakly coupled candidate particle that could satisfy these constraints. Photons or relativistic (massless) neutrinos can be excluded because of their too small interaction cross sections or, equivalently, too long mean free times. On the other hand, gravitons are expected
to have a mean free time~\cite{Hawking:1966qi}
\begin{equation}
\tau_G = \frac{1}{16 \pi G_\text{N} \eta}.
\label{eq26}
\end{equation}
One can solve eqs.\ \eqref{eq24} and \eqref{eq25} for $\eta$ and $\tau_G$~\cite{Weinberg1972}, and one finds
\begin{equation}
\frac{4\pi G_\text{N} D}{3 H^3} = \frac{4\pi G_\text{N} \eta \sigma}{3 H } =  \sigma \sqrt{\frac{c_\eta\, \epsilon_G}{24\, \rho_c}}  \, .
\label{eq27}
\end{equation}
For an accelerating expansion, one would have to require $\sigma \gtrsim 10$, $c_\eta \approx 1$, and most importantly, a fractional energy density of the gravitational radiation $\Omega_G = \epsilon_G/\rho_c$ not too far from one. This would also satisfy the consistency condition $\tau_G H \lesssim 1$. The purpose of the above comment is not to argue that a graviton gas of such high energy density, interacting with dark matter, can provide a phenomenologically viable component of the dark sector (in any case, this would only seem plausible if such a component plays essentially the role of $\Omega_\Lambda$ in the standard concordance model). Rather, we sketch this scenario only to illustrate with an example how a specific particle content of the dark sector affects its material properties, and how these properties may impact the large-scale dynamics of the universe, or can be constrained by it.

In summary, the main result of this letter is the identification of a dissipative term $D$ in the cosmological evolution \eqref{eq17} of the average energy density. This term arises from the backreaction of fluid velocity fluctuations, depends on shear viscosity, and may affect the expansion history of the universe. If sufficiently large, it could lead to an accelerating  cosmological expansion without assuming negative effective pressure. Since the shear viscous and bulk viscous fluctuations measured by $D$ are expected to take significant values only during the epoch of structure formation, this would also provide a natural explanation for why an accelerated cosmological expansion occurs only at late times in the history of the universe. Irrespectively of the size of $D$, we emphasize that dissipative phenomena are ubiquitous in nature, and that eqs.~\eqref{eq18} and~\eqref{eq22} for $D$ provide a novel and more comprehensive framework to account for them in discussions of the cosmological expansion. At least in principle, eqs.~\eqref{eq18} or ~\eqref{eq22} can be calculated also for non-equilibrium scenarios, which is of interest since it is a priori unclear to what extent the dark sector is equilibrated. Also, it is conceivable that contributions to $D$ arise from sources not discussed so far. For instance, an effective viscosity may also arise on large length scales from a coarse-grained  description of fluctuations in the cosmological fluid \cite{Calzetta:1991xe,Baumann:2010tm}. Or, at least in principle, a contribution to $D$ could also arise from the contraction of the fluid against local pressure gradients that might be induced by gravitational collapse (third term in eq.~\eqref{eq18}). In view of these many physics effects, we hope that the
results derived in this letter will help to better constrain the role of dissipation in cosmology, and the material properties that may give rise to it.

\paragraph*{Acknowledgments}
We acknowledge useful discussions with D.~Blas, M.~Garny, M.~Pietroni and S.~Sibiryakov.


\end{document}